\documentclass[preprint,showpacs,amsmath,amssymb,floatfix,pre]{revtex4}
\usepackage{graphicx}
\begin{document}
\title{The magnetic susceptibility on the transverse antiferromagnetic Ising model: Analysis of the reentrant behaviour}

\author{Minos A. Neto}
\email{minos@pq.cnpq.br}
\author{J. Ricardo de Sousa}
\author{Igor T. Padilha}
\author{Octavio R. Salmon}
\author{J. Roberto Viana}
\affiliation{Departamento de F\'{\i}sica, Universidade Federal do Amazonas, 3000, Japiim,
69077-000, Manaus-AM, Brazil}
\author{F. Din\'{o}la Neto}
\affiliation{Cavendish Laboratory, University of Cambridge; J. J. Thomson Avenue, Cambridge, CB3 0HE, England.}
\date{\today}

\begin{abstract}
\textbf{ABSTRACT}

We study the three-dimensional antiferromagnetic Ising model in both uniform longitudinal ($H$) and transverse ($\Omega $) magnetic fields 
by using the effective-field theory with finite cluster $N=1$ spin (EFT-1). We analyzed the behavior of the magnetic susceptibility 
to investigate the reentrant phenomena we have seen the same phase diagram previously obtained in another papers. Our results shows the 
presence of two divergences in the susceptibility that indicates the existence of a reentrant behaviour.

\end{abstract}

\maketitle

\section{Introduction\protect\nolinebreak}

The Ising antiferromagnet (AF) in transverse and longitudinal magnetic fields divided into two equivalent interpenetrating sublattices $A$ 
and $B$, which is described by the following Hamiltonian

\begin{equation}
H=J\sum\limits_{<i,j>}\sigma _{i}^{z}\sigma_{j}^{z}-H\sum\limits_{i}\sigma _{i}^{z}-\Omega \sum\limits_{i}\sigma_{i}^{x},
\label{1}
\end{equation}%
where $J$ is a positive coupling constant, which stands for the antiferromagnetic interaction between the magnetic moments (electrical origin), 
$\langle i,j\rangle$ represents the sum over all pairs of nearest-neighbour spins on a cubic lattice, $\sigma_{i}^{\nu }$ is the $\nu (=x,z)$ component, 
at site $i$, of the spin-$1/2$ Pauli operator, $\Omega$ and $H$ stand for the transverse and longitudinal magnetic fields, respectively.

The operators $\sigma_{i}^{x}$ and $\sigma_{i}^{z}$  do not commute, thus the field $(\Omega)$, which is transverse to the spin direction causes quantum 
spin fluctuations and quantum tunnelling between the spin-up and spin-down eigenstates of $\sigma_{i}^{z}$. The fluctuations reduce the critical temperature 
$T_{c}$ below which the spins maintain a long-range order. The critical temperature vanishes at a critical field $\Omega_{c}$, and a quantum phase transition 
occurs between the antiferromagnetic and paramagnetic (P) quantum state.

The magnetic susceptibility measures the capacity of a material to magnetise under the action of an external magnetic field. This property has been 
studied extensively in recent years. Some new results for the magnetic susceptibility for crystal fields and Kondo effect were obtained by Desgranges 
\cite{desgranges2015}. Kashif, \textit{et al.} \cite{kashif2008} have studied the effects of copper addition on density and magnetic 
susceptibility of lithium borate glasses: $\left(100-x\right)(Li_{2}O\cdot2B_{2}O_{3})\cdot xCuO$, where $x=0$, $5$, $10$, $15$, $20$ and $25$ mol$\%$. 
At low temperatures, the dc magnetic susceptibility measurements for the heavy fermion $CeCu_{6-x} Au_{x}$ show an anomalous behaviour for the quantum 
critical concentration $x=0.1$: The temperature dependence near to the quantum phase transition $\chi^{-1} \sim \theta +T^{\alpha}$ 
shows a coefficient $\alpha=0.8$ \cite{tomanic2008}. These results have been confirmed in some literatures \cite{schroder1998,schroder2000}. 

Recently, an intrinsic magnetic susceptibility of height purified single-wall carbon nanotubes \cite{nakai2015}. The effect of fast neutron irradiation 
on the magnetic susceptibility and magneto-resistance of $Si$ whiskers with impurity concentration in metal-insulator transition they were observed by 
Druzhinin, \textit{et al.} \cite{druzhinin2015}. Inelastic and quasi-elastic excitations in the neutron scattering on the compounds $CeCo_{1-x}Cu_{x}Al_{4}$ 
$(x=0$, $0.05$, $0.1)$ have been studied the antiferromagnetic ordering temperature, where the crystal field manifestation in inelastic neutron scattering, 
magnetic susceptibility and specific heat can be observed on the compounds $CeCoAl_{4}$ \cite{tolinski2013}. Experimentally, another interesting analysis 
is the magnetism of linear chains antiferromagnetic at high temperatures by study of the dynamics susceptibility \cite{nijhof1986} of the hydrated compounds 
$RbCoCl_{3}\cdot2H_{2}O$.

For these several years \cite{frus_comp}, magnetic phenomena with frustration or competing interations have been studied, because it is considered to be one 
of the most essential mechanisms for spin glass (SG) phenomena \cite{kitatani1986}. One of the characteristics that appear in this type of phenomena is the 
reentrant behaviour. This phenomenon can be seen in amorphous materials such as e.g. $\left(Fe_{1-x}Mn_{x}\right)_{75}P_{16}B_{6}Al_{3}$, which displays the 
reentrance for several concentration $x$ \cite{mirebeau1990}. 

More recently, a reentrant spin glass behaviour in CE-type AFM $Pr_{0.5}Ca_{0.5}MnO_{3}$ manganite in the sequence of multiple magnetic transitions it was 
observed \cite{cao2006}, showing a transition: $P \rightarrow F \rightarrow AF \rightarrow SG$. The prove of the existence of the SG phase, in this polycrystalline 
material, can be seen in behaviour of the susceptibility versus temperature. Another SG material that has the reentrant behaviour is the 
$IrSr_{2}Sm_{1.15}Ce_{0.85}Cu_{2.175}O_{10}$ \cite{colman2012}. The reentrant behaviour, in this compounds, it can be seen via magnetic property measurements of AC 
and DC susceptibility.

The pure transverse Ising model (TIM) has been used to describe a variety of physical systems, for example, correlation functions for lattice gauge theories with 
action Boltzmann factor \cite{barreto2015},  thermodynamical properties of the mixed with four-spin interactions \cite{ghliyem2014}, critical behavior in two dimensional 
of the fidelity susceptibility \cite{nishiyama2013}, effect of surface dilution \cite{kaneyoshi2012}, description ferromagnetism in a transverse Ising antiferromagnet 
\cite{kaneyoshi2016}, reentrant phenomenona in nanosystems \cite{kaneyoshi2013} and effects of the randomly in thin film \cite{umit2013}. 

Using EFT-1 with finite cluster N=1 spin, in the present paper, we investigate the magnetic susceptibility analysis on the transverse of the Ising antiferromagnet 
in both external longitudinal and transverse fields. This work is organized as follows: In Sec. II we outline the formalism and its application to the transverse Ising 
antiferromagnetic longitudinal magnetic field; in Sec. III we discuss the results; and finally, in Sec. IV we present our conclusions.

\section{Formalism}

It was originally introduced by de Gennes \cite{deGennes} as a pseudospin model for hydrogen-bonded ferroelectric such as KH$_{2}$PO$_{4}$ \cite{blinc,elliot}. 
Theoretically, various techniques have been used in the transverse Ising model (TIM) as: renormalization group (RG) method \cite{4}, cluster variation method (CVM) 
\cite{8}, mean field theory (MFT) \cite{9}, pair approximation (PA) \cite{10}, Monte Carlo (MC) simulations \cite{11} and effective-field theory (EFT) 
\cite{5,6,7,hk,denise2013,denise2012,antonia14} (we'll deal with here in this work).

In what the  ground-state of the model concerns (\ref{1}), it  is characterized by an antiparallel spin alignment in the horizontal and vertical directions, thus 
it exhibits N\'{e}el ordering within the initial sublattices $A$ and $B$. This is the AF state. This state must not be confused with the superantiferromagnetic (SAF) 
state, because the values of the longitudinal field  $H^{AF}\neq H^{SAF}$ are different in the ground state. On the other hand, the phase diagram exhibits interesting 
properties due to  the competition between the antiferromagnetic exchange interaction and the longitudinal field $H$. Particularly, the model (\ref{1}) has an AF 
phase in the presence of a field, where the temperature of the transition decreases as the intensity of the fields increases, so at $T=0$, a second-order phase 
transition occurs at critical values of $H_{c}$ and $\Omega _{c}$.

In order to treat the model (\ref{1}) on a simple cubic lattice through the EFT-1 approach, we consider a simple example of cluster on a lattice consisting of a central 
spin surrounded by $z$ spins. Then, these $z$ spins are substituted by an effective field, which is proportional to  the thermal average of the central spin. Accordingly, 
the Hamiltonian for this cluster is given by

\begin{equation}
\mathcal{H}_{1A}=\left( J\overset{z}{\sum\limits_{\delta }}\sigma
_{(1+\delta)B}^{z}-H\right) \sigma _{1A}^{z}-\Omega \sigma _{1A}^{x},
\label{2}
\end{equation}%
and 
\begin{equation}
\mathcal{H}_{1B}=\left( J\overset{z}{\sum\limits_{\delta }}\sigma
_{(1+\delta )A}^{z}-H\right) \sigma _{1B}^{z}-\Omega \sigma _{1B}^{x}.
\label{3}
\end{equation}%
From the Hamiltonians (\ref{2}) and (\ref{3}), we obtain the average magnetizations in sublattices $A$, $m_{A}=\langle \sigma _{1A}^{z}\rangle$, and $B$, $m_{B}=\langle 
\sigma _{1B}^{z}\rangle$, using the approximate Callen-Suzuki \cite{barreto}, which are given by

\begin{equation}
m_{A}=\left\langle \frac{H-a_{1A}}{\sqrt{(H-a_{1A})^{2}+\Omega ^{2}}}\tanh
\beta \sqrt{(H-a_{1A})^{2}+\Omega ^{2}}\right\rangle,
\label{4}
\end{equation}%
and 
\begin{equation}
m_{B}=\left\langle \frac{H-a_{1B}}{\sqrt{(H-a_{1B})^{2}+\Omega ^{2}}}\tanh
\beta \sqrt{(H-a_{1B})^{2}+\Omega ^{2}}\right\rangle,
\label{5}
\end{equation}%
\textit{where $a_{1A}=J\overset{z}{\sum\limits_{\delta }}\sigma_{(1+\delta )B}^{z}$ and $a_{1B}=J\overset{z}{\sum\limits_{\delta }}\sigma _{(1+\delta )A}^{z}$.}

Now, after using  the identity $\exp (\alpha D_{x})F(x)=F(x+a)$ (where $D_{x}=\frac{\partial }{\partial x}$ is the differential operator) and the van der Waerden 
identity for the two-state spin system (i. e., $\exp (a\sigma_{i}^{z})=\cosh (a)+\sigma _{i}^{z}\sinh (a)$) the Eqs. (\ref{4}) and (\ref{5}) are rewritten as follows

\begin{equation}
m_{A}=\left\langle \prod_{\delta\neq 0}^{z}(\alpha
_{x}+\sigma _{(1+\delta )B}^{z}\beta _{x})\right\rangle \left.
F(x)\right\vert _{x=0},
\label{6}
\end{equation}%
and
\begin{equation}
m_{B}=\left\langle \prod_{\delta\neq 0}^{z}(\alpha
_{x}+\sigma _{(1+\delta )A}^{z}\beta _{x})\right\rangle \left.
F(x)\right\vert _{x=0},
\label{7}
\end{equation}%
with 
\begin{equation}
F(x)=\frac{H-x}{\sqrt{(H-x)^{2}+\Omega ^{2}}}\tanh \beta \sqrt{%
(H-x)^{2}+\Omega ^{2}},
\label{8}
\end{equation}%
where $\beta _{x}=\sinh (JD_{x})$ and $\alpha _{x}=\cosh (JD_{x})$. The Eqs. (\ref{6}) and (\ref{7}) are rewritten in terms of multiple spin correlation 
functions. We remark that it is very difficult  to address all the correlations between the spins in Eqs. (\ref{6}) and (\ref{7}). So, in this work we resort  to decouple the 
right-hand sides of Eqs. (\ref{6}) and (\ref{7}), in such a way that

\begin{equation}
\left\langle \sigma_{iA}^{z} \sigma_{jB}^{z} \ldots \sigma_{lA}^{z}\right\rangle \backsimeq m_{A} m_{B}\ldots m_{A}, 
\label{8a}
\end{equation}%
where $i\neq j \neq \ldots \neq l$ and $m_{\mu}=\left\langle \sigma_{i\mu}^{z} \right\rangle $ $(\mu=A,B)$. Although the approximation (\ref{8a}) disregards correlations 
between different spins, it treats relations such as $\left\langle \left( \sigma_{i\mu}^{z} \right)^{2} \right\rangle =1$ exactly, whereas in the usual MFT all the self- 
and multi-spin correlations are neglected. Then, by the use of the approximation (\ref{8a}), the Eqs. (\ref{6}) and (\ref{7}) are rewritten as follows

\begin{equation}
m_{A}=\sum_{p=0}^{z} A_{p}(T_{N},h,\delta)m_{B}^{p},
\label{9}
\end{equation}%
and 
\begin{equation}
m_{B}=\sum_{p=0}^{z} A_{p}(T_{N},h,\delta)m_{A}^{p},
\label{10}
\end{equation}%
with

\begin{equation}
A_{p}(T_{N},h,\delta)=\frac{z!}{p!(z-p)!}\alpha _{x}^{z-p}\beta _{x}^{p}\left.
F(x)\right\vert _{x=0},
\label{11}
\end{equation}%
where the coefficients $A_{p}(T_{N},h,\delta)$ are obtained by means of the mathematical property $\exp (\alpha D_{x})F(x)=F(x+a)$.

Close to the critical point, and in terms of the uniform $m=\frac{1}{2}(m_{A}+m_{B})$ and staggered $m_{s}=\frac{1}{2}(m_{A}-m_{B})$ magnetizations, we have $m_{s}\rightarrow 0$ 
and $m\rightarrow m_{0}$, thus, the expansion  of the sublattice magnetization $m_{A}$ (up to linear order in $m_{s}$) is given by

\begin{equation}
m_{A}=\omega_{0}(T_{N},h,\delta,m_{0})+\omega_{1}(T_{N},h,\delta,m_{0})m_{s},
\label{12}
\end{equation}%
with 
\begin{equation}
\omega_{0}(T_{N},h,\delta,m_{0})=\sum_{p=0}^{z}A_{p}(T_{N},h,\delta)m_{0}^{p},
\label{13}
\end{equation}%
and 
\begin{equation}
\omega_{1}(T_{N},h,\delta ,m_{0})=-\sum_{p=0}^{z} pA_{p}(T_{N},h,\delta)m_{0}^{p-1}. 
\label{14}
\end{equation}

On the other hand, due to the fact that we only observed  second-order phase  transitions, we analysed only the Eqs. (\ref{13}) and (\ref{14}) in the limit of $m_{\mathbf{s}}\rightarrow 0$, 
so as to obtain the phase diagram. So we could locate the second-order line using the fact that $m_{A}=m_{0}+m_{s}$ in Eq. (\ref{12}), thus:

\begin{equation}
\omega_{0}(T_{N},h,\delta ,m_{0})=m_{0}
\label{15}
\end{equation}%
and
\begin{equation}
\omega_{1}(T_{N},h,\delta,m_{0})=1,
\label{16}
\end{equation}%
at the critical point in which $m_{s}=0$, $\delta=\Omega/J$ and $h=H/J$.

Therefore, we could find a solution for the second-order critical frontier, being $m_{s}$ the order parameter, which is suitable for describing the phase transition of the model (\ref{1}). 
It is important to mention that the magnetizations of the two sublattices are not equal for $m_{s}\neq 0$, and the system is in the antiferromagnetic phase. Nervertheless, the magnetizations 
of the two sublattices are equal for $m_{s}=0$, so the system is in the saturated P phase.

The initial magnetic susceptibility $\chi_{s}$ is calculated by the derivative of the magnetization function $m_{s}(H)$ given by the subtraction of the equations (\ref{9}) 
and (\ref{10}), i. e.,

\begin{equation}
\chi_{s}=\left(\frac{\partial m_{s}}{\partial H} \right)_{H=0}=\frac{\Theta}{1+\Delta},
\label{17}
\end{equation}%
where $\Theta(m,m_{s})$ and $\Delta(m,m_{s})$.

\section{Results and Discussions}

\begin{figure}[htbp]
\centering
\includegraphics[width=7.0cm,height=7.0cm]{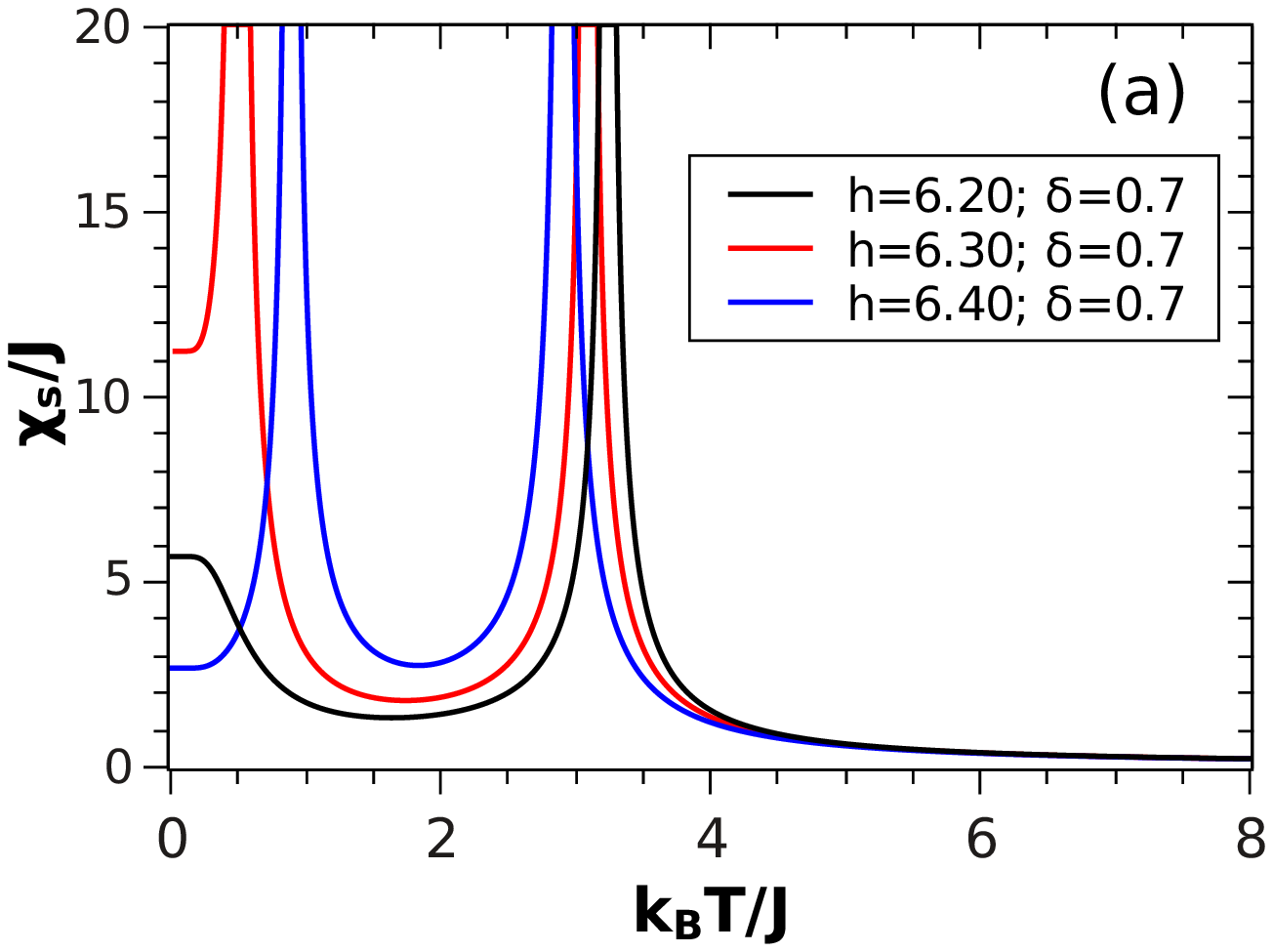}
\hspace{0.4cm}
\includegraphics[width=7.0cm,height=7.0cm]{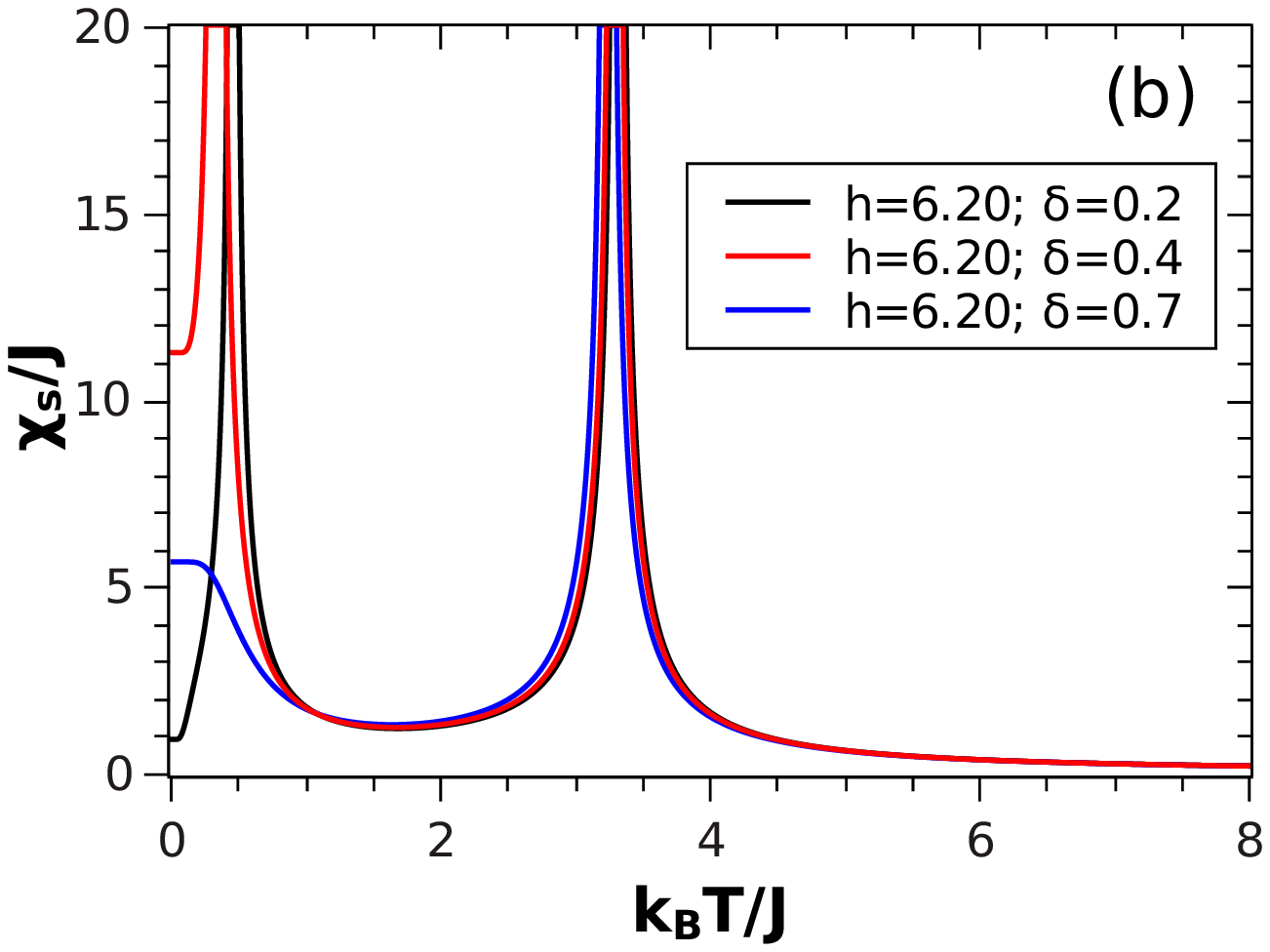}
\hspace{0.4cm}
\includegraphics[width=7.0cm,height=7.0cm]{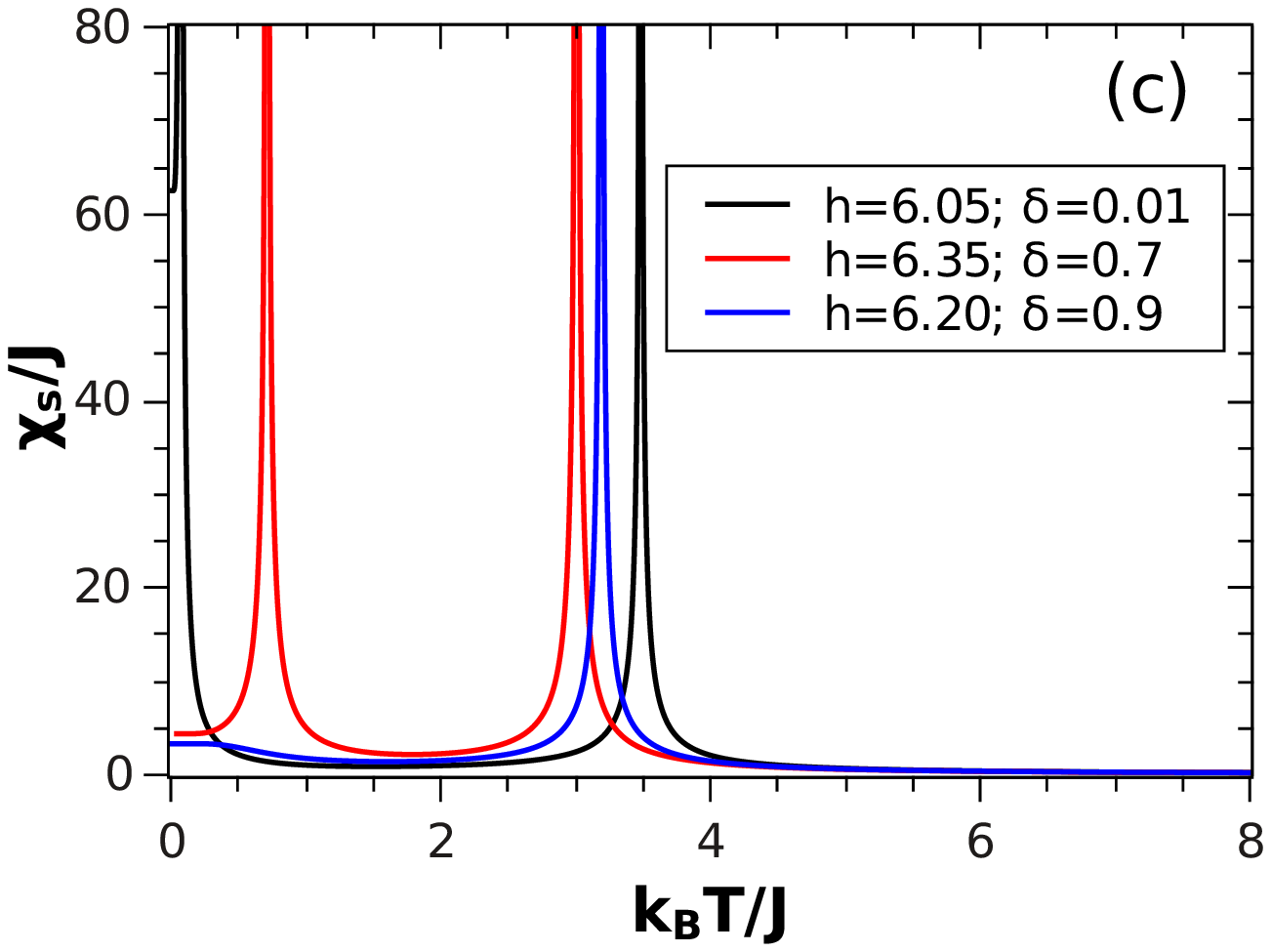}
\hspace{0.4cm}
\includegraphics[width=7.0cm,height=7.0cm]{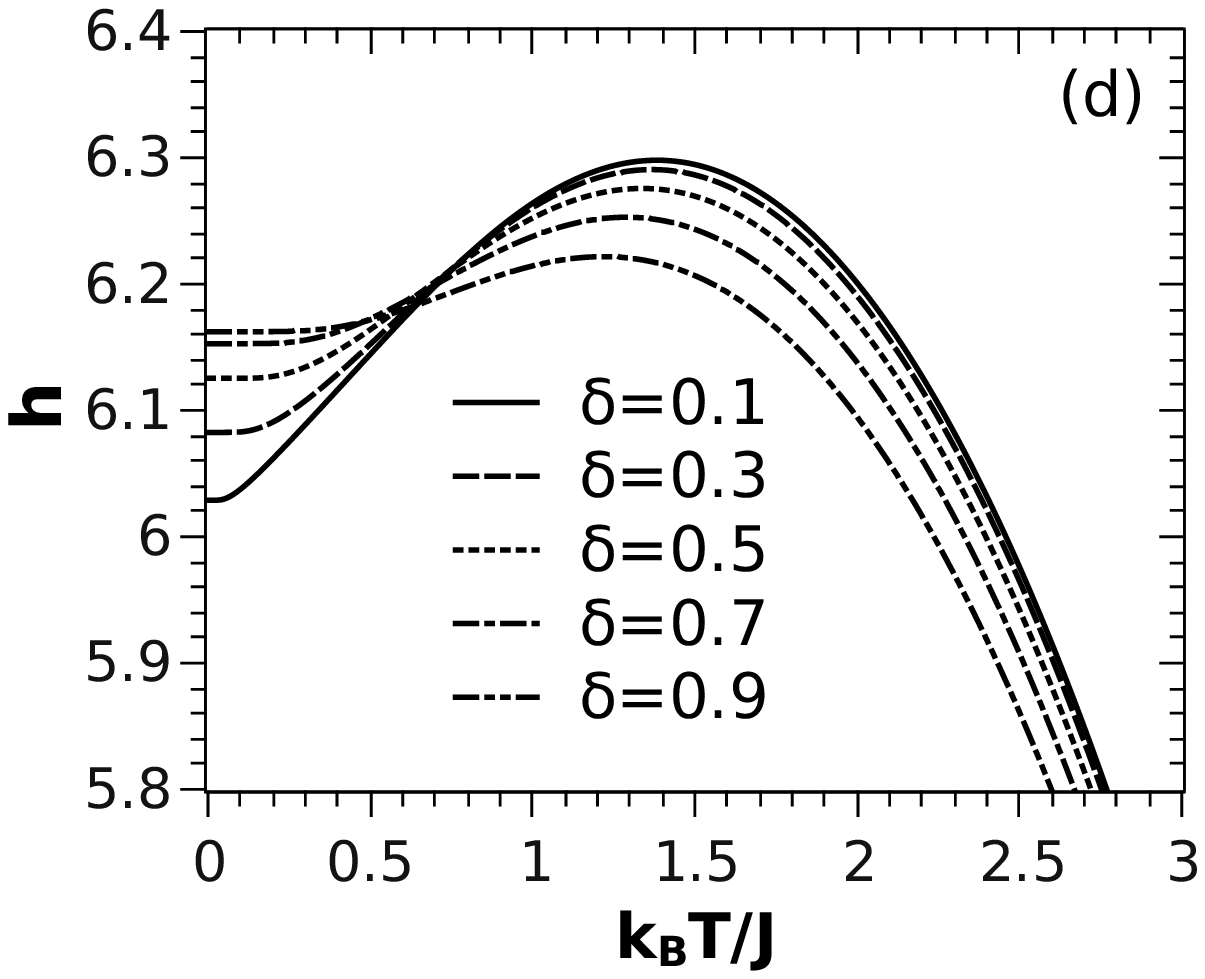}

\caption{Dependence of the susceptibility $\chi_{s}/J_{y}$, as a function of the temperature. In the Fig. (\ref{susc_s1}.a) we show this dependence by setting 
the value of the transverse magnetic field $\delta=0.7$ and we vary the longitudinal magnetic field. In the Fig. (\ref{susc_s1}.b) we show this dependence by 
setting the value of the longitudinal magnetic field $\delta=6.20$ and we vary the transverse magnetic field. In the Fig. (\ref{susc_s1}.c), we vary both the 
fields. In the Fig. (\ref{susc_s1}.d), we reproduce the $h-T$ phase diagram obtained by Neto and de Sousa \cite{minos2013} we show for the reentrant region.} 
\label{susc_s1}
\end{figure}

In Figs. \ref{susc_s1}. (a-c), we show the behaviour of the magnetic susceptibility as a function of temperature by varying the values of the magnetic fields $h$ 
and $\delta$. The numerical determination of the phase boundary (second-order phase transition) is obtained  by solving simultaneously the set of Eqs. (\ref{15}) 
and (\ref{17}). In Figs. (\ref{susc_s1}.a) and (\ref{susc_s1}.b). We set the values of the magnetic fields $\delta$ (fixed $h$) and $h$ (fixed $\delta$), respectively, 
where we can see clearly the existence of the reentrant phase transition ($P\rightarrow AF\rightarrow P$). It is a well-known example is the cuprate material, 
$IrSr_{2}Sm_{1.15}Ce_{0.85}Cu_{2.175}O_{10}$, shows in measures of DC-susceptibility an evidence of complex set of magnetic transitions upon cooling that are 
characteristic of a reentrant SG ground-state.

In Fig. \ref{susc_s1}.(a), for $h=6.20$ and $\delta=0.7$, the phase transition occurs at only one point (divergence of $\chi_{s}$) in $T_{N}^{(1)}=3.230$. However, 
inside the reentrant region when we increased the value of $h$, we find another transition peak. In the same figure, we find that for $h=6.30$ and $\delta=0.7$ we 
have $T_{N}^{(1)}=0.486$ and $T_{N}^{(2)}=3.082$, and for $h=6.40$ and $\delta=0.7$ we have $T_{N}^{(1)}=0.486$ and $T_{N}^{(2)}=3.082$, respectively. In the limit 
of high temperature the magnetic susceptibility as a function of the temperature presents a behaviour kind $1/T$.  

In Fig. \ref{susc_s1}. (b), the same behavior appears when we fix the value of $h$ and increase the value of $\delta$. For case $h=6.20$ and $\delta=0.2$, the phase 
transition occurs at two one point in $T_{N}^{(1)}=0.452$ and $T_{N}^{(2)}=3.303$, and for case $h=6.20$ and $\delta=0.4$, the phase transition occurs in the points 
$T_{N}^{(1)}=0.326$ and $T_{N}^{(2)}=3.281$. We note that, fixing the value of $h$ and varying $\delta$ the values of the temperatures of transition decreases. A 
possible, degeneracy and increase in the energy of the ground-state these regions may occur in magnetic models with competitive interactions are sensitive the these 
perturbation in low-temperature.

The two transition points can also be seen in Fig. \ref{susc_s1}. (c), where we observe three behavior of the susceptibility as a function of temperature with some 
values of $h$ and $\delta$. For case $h=6.05$ and $\delta=0.01$, the phase transition occurs at two one point in $T_{N}^{(1)}=0.057$ and $T_{N}^{(2)}=3.473$, the 
case $h=6.35$ and $\delta=0.7$, the phase transition occurs at two one point in $T_{N}^{(1)}=0.7$ and $T_{N}^{(2)}=2.995$ and the case $h=6.20$ and $\delta=0.9$, the 
phase transition occurs in the one point $T_{N}^{(1)}=3.179$. For the occurrence of these reentrant phenomenon in a certain appropriate range of $h$ and $\delta$, 
explanation of mechanisms has been given by Neto and de Sousa \cite{minos2013}. In this work the phase diagram in the $h-T$ plane is presented with selected values 
of $\delta$ that show the presence of reentrant behaviour in the region of low temperature, see Fig. (\ref{susc_s1}.d). 

\section{Conclusions}

These magnetic phenomena in which frustration and competitive interactions are involved, have been studied due to the fact that they are  considered one of the most 
essential mechanism for spins glass phenomena and more recently in behaviours of multiple reentrant glass transitions in confined hard-sphere glasses \cite{mandal2014}. 
This phenomenon appears in glass transition in a colloid polymer mixture with depletion attractions \cite{eckert2002} showing that in the introduction in the short-range 
attraction to a colloid suspension of nearly hard spheres by addition of a free polymer produces this phenomenon. 

In both cases it appears the two transition points near the second transition point the derivative $d\chi_{s}/d T>0$ is positive and at high-temperature the 
entropy is the predominant factor and the system is then in the disordered P phase but with an AF bias due the applied fields. This is a good model that could represent 
multiple magnetic transitions as it is seen experimentally in some materials \cite{cao2006,colman2012}. As expected, $\chi_{s}$ increases rapidly with increasing temperature 
and diverges at the critical point $T_{c}$ and and behavior is presented as $1/T$ in the limit of high-temperature, as expected.

In this work we analyze the reentrant behavior on the magnetic susceptibility on the region of the reentrant phenomenon using the EFT-1. Our results are consistent with second-order 
transitions from the AF ordered to the P disordered phase at zero transverse field. Furthermore, the investigations of this model with quenched disorder is 
expected to show many characteristic phenomena, as already analysed in models of metal-insulator transition \cite{bozin2014}. This will be discussed in future work.

\textbf{ACKNOWLEDGEMENT}

This work was partially supported by FAPEAM and CNPq (Brazilian Research Agency).

\vspace{0.4cm}

\end{document}